\documentclass{eas}
\usepackage{graphicx}
\usepackage{natbib}
\begin{document}

\title{The Variability Processing and Analysis of the Gaia mission} 
\runningtitle{Eyer \etal: The Gaia Variability Processing and Analysis}
\author{Laurent Eyer}\address{Department of Astronomy, University of Geneva, CH 1290 Versoix  (\email{Laurent.Eyer@unige.ch})}
\author{Dafydd Wyn Evans}\address{Institute of Astronomy, University of Cambridge, CB3 0HA, UK}
\author{Nami Mowlavi}\sameaddress{1}
\author{Alessandro Lanzafame}\address{Dipartimento di Fisica e Astronomia, Universit\`a di Catania, Via S. Sofia 78, Catania, Italy}
\author{Jan Cuypers}\address{Royal Observatory of Belgium, Ringlaan 3, Brussels, Belgium}
\author{Joris De Ridder}\address{Instituut voor Sterrenkunde, KU Leuven, Celestijnenlaan 200D, B-3001 Leuven}
\author{Luis Sarro}\address{Departamento de Inteligencia Artificial, UNED, Juan del Rosal, 16, 28040 Madrid, Spain}
\author{Gisella Clementini}\address{INAF Osservatorio Astronomico di Bologna, Via Ranzani 1, 40127 Bologna, Italy}
\author{Leanne Guy}\sameaddress{1}
\author{Berry Holl}\sameaddress{1}
\author{Diego Ordonez}\sameaddress{1}
\author{Krzysztof Nienartowicz}\sameaddress{1}
\author{Isabelle Lecoeur-Taibi}\sameaddress{1}

\begin{abstract}
We present the variability processing and analysis that is foreseen for the Gaia mission within Coordination Unit~7 (CU7) of the Gaia Data Processing and Analysis Consortium (DPAC). A top level description of the tasks is given.
\end{abstract}
\maketitle
\section{Introduction}
Variable phenomena are common in the Universe and multi-epoch surveys are producing very interesting scientific results
\citep[see][]{EyerMowlavi2008}. Among all the numerous existing surveys, Gaia is really prominent and will have
a tremendous impact on Time Domain Astronomy \cite[see for example][]{Eyeretal2012, Eyeretal2013}.
Within the Gaia Data Processing and Analysis Consortium \cite[DPAC, see][]{Mignardetal2008}, one Coordination Unit (CU7) is dedicated to the thematics of variability.  The objective of CU7 is to populate the Gaia catalogue with properties of sources having photometric and/or spectral variability.

CU7 is composed of about 70  researchers, postdocs, students, and computers scientists who are distributed among about 20 countries.

\section{Variability analysis Work Breakdown}

\begin{figure}
\includegraphics[width=9cm,angle=-90]{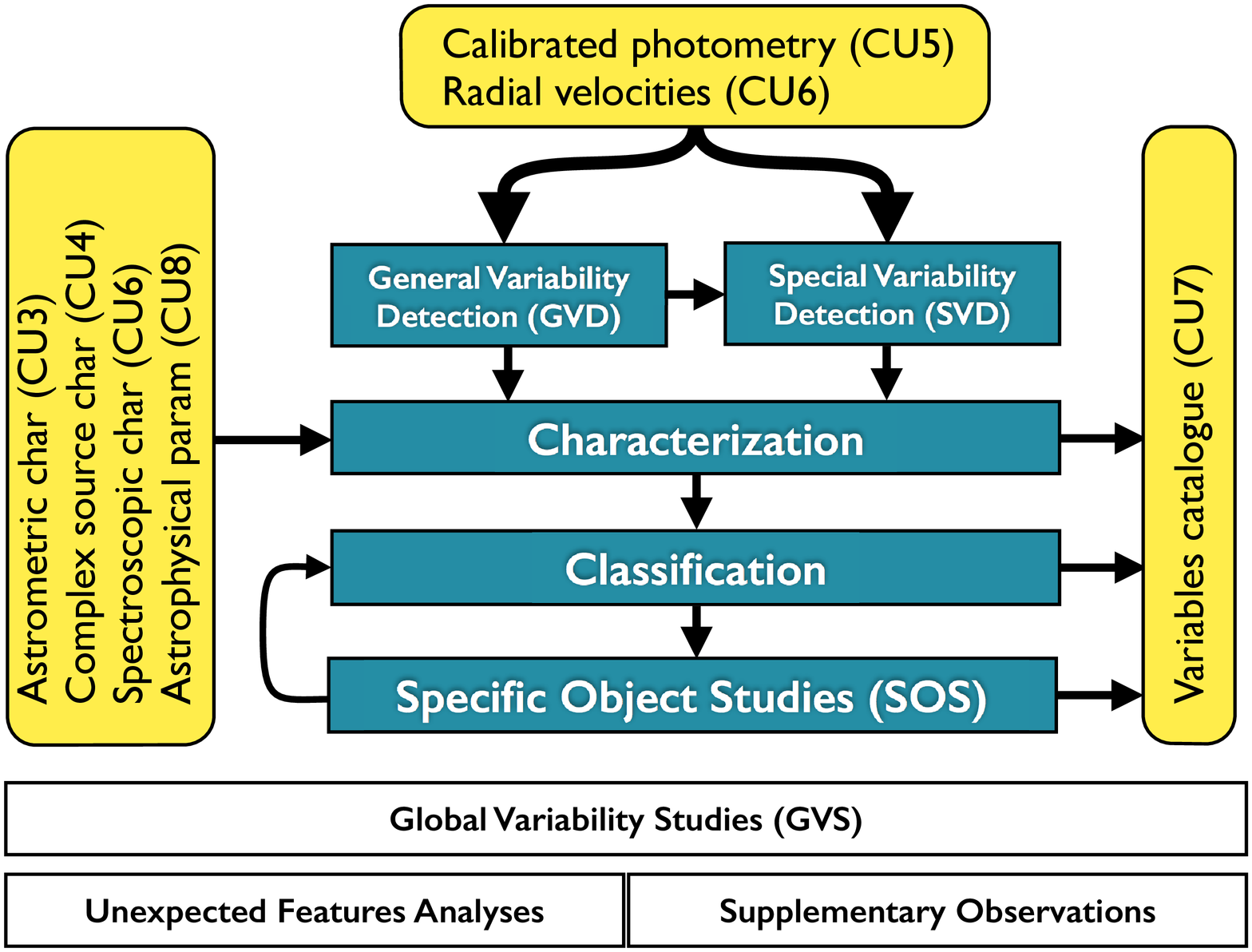} \caption{\label{FIGCU7WPs} Schematic Gaia Variability Processing and Analysis Work Breakdown.}
\end{figure}

The top-level work-breakdown is represented in Fig.~\ref{FIGCU7WPs}. We describe briefly the different tasks hereafter.

{\bf Variability Detection:}
Variability detection comprises both statistical tests for the general detection of variability as well as 
specific algorithms that leverage knowledge of a specific variability behaviour to achieve a higher detection efficiency.
These latter algorithms ensure variability detection in cases where general variability tests are not sufficient and in addition derive some parameters  
specific to the type of variability. The cases treated by such specific algorithms are: short time-scales \citep{Varadi2011}, small amplitude periodic variables, exo-planetary transits \citep{Tingley2011, Dzigan2012, Dzigan2013}, solar (magnetic) variables \citep{Distefanoetal2012}.

 {\bf Characterization:} The variability behaviour is characterized using a classical Fourier decomposition approach; the goal being to find the simplest possible description of the observed variability.  Each source is scanned for periodicity and the most significant period, if significant enough, is retained and a  Fourier series model fit. If the model residuals still show variability, according to the criteria defined above, an additional period search will be initiated to improve the final model.  If no significant period is found, the final model will be a polynomial only and the source is  designated as non-periodically variable. The parameters of the model and all computed statistical parameters are used to construct the attributes for Classification. A lot of attention was given to optimizing the period search computation to find the best compromise between speed and efficiency. Significant theoretical research went into the search for the most appropriate false-alarm probability for periodicity for these data sets. 

{\bf Classification:} Once variables are identified and characterised, they go through different classification methods: Supervised, clustering methods and extractors (microlensing, transients, massive companions in binary systems). The result is a number of groups containing objects with similar variability behaviour (and most probably similar physical properties).
Methods for extractors and supervised algorithms have been tested on Hipparcos data \cite{Dubathetal2011}, \cite{Rimoldinietal2012}.

{\bf Specific Object Studies:}
Once the variability type of a given source is assigned by Classification, specific algorithms are applied to validate it and to compute additional attributes.
Some examples are:
the Paczy\'nski parameters of microlensing events;
a sub-classification of eclipsing binaries;
the computation of flare occurrence and magnitude for flaring stars,
 of rotation period and light-curve amplitude for rotation-induced variables,
 or of burst and outburst parameters for transients such as Be stars;
the characterisation of pulsating stars such as long period variables, Cepheids, RR~Lyrae, or of very short period variables like AM~CVn or ZZ~Ceti stars.

{\bf Global Variability Studies:} Subsequent to the source-by-source
processing, all possible information about Gaia variables is available
in the Variability Database. Several catalogue-level results such as
the period distribution for all Cepheids, the colour-magnitude or HR
diagram with iso-contour of variability amplitude, etc. are derived.
The statistical properties of the Gaia Variability Database are
studied compared to those of similar surveys in order to assess the
quality of the Variability Database and to identify possible errors or
problems in the data processing.  Tools to aggregate, evaluate and
visualize the database content have been developed.


{\bf Supplementary Observations:} Finally, there are cases where supplementary observations, either from ground-based observatories, or from other satellites, 
will be required to complete our knowledge about the Gaia sources. The use of these observations is to help the development of the data processing or to test the output of the Gaia analysis. A network of small medium-sized telescopes was put in place and supplementary observations (photometry and/or spectroscopy) were collected for samples of Cepheids, RR Lyrae stars, Miras, short-period variables and Be stars.
The network of telescope is also actively participating in the validation of the Gaia Alerts stream.

\section{Geneva Data Processing Centre}

The software modules delivered by CU7 are described in Section 2, the releases of these modules are integrated at DPCG, the Geneva Data Processing Centre. The software once integrated receives the name of VariPipe. During every processing cycle in Gaia a new release of VariPipe is produced and deployed in the processing cluster in Geneva to obtain the scientific results out of the Gaia data \citep[see][]{Nienartowiczetal2014}. The DPCG is also the interface with the other data processing centres of the consortium and the Science Operations Centre. The data produced by the upstream CUs is received and prepared for its analysis, the software chains are run on the data and the produced results are prepared to be integrated in the Gaia catalogue. DPCG provides tools to analyze the input data as well as the results to ensure the quality of the delivered products to the Gaia catalogue.

\section{Releases}
A first global release of all variable sources detected among the one billion observed sources is foreseen in the 4th release (currently planned for 2018/2019), the final release might come in 2022. However the goal of CU7 is to release at earlier dates specific groups of variables, once the estimations of completeness and contamination are thought to be reliable enough and reach acceptable levels.



\begin{thebibliography}{99}
\bibitem[Distefano et al.(2012)]{Distefanoetal2012} Distefano, E., 
Lanzafame, A.~C., Lanza, A.~F., et al.\ 2012, MNRAS, 421, 2774 
\bibitem[Dubath et al.(2011)]{Dubathetal2011} Dubath, P., Rimoldini, 
L., S{\"u}veges, M., et al.\ 2011, MNRAS, 414, 2602
\bibitem[Dzigan \& Zucker(2012)]{Dzigan2012} Dzigan, Y., \& Zucker, S.\ 2012, ApJ, 753, LL1 
\bibitem[Dzigan \& Zucker(2013)]{Dzigan2013} Dzigan, Y., \& Zucker, S.\ 2013, MNRAS, 428, 3641 
\bibitem[Eyer \& Mowlavi(2008)]{EyerMowlavi2008} Eyer, L., \& Mowlavi, N.\ 2008, Journal of Physics Conference Series, 118, 012010
\bibitem[Eyer et al.(2012)]{Eyeretal2012} Eyer, L., Palaversa, L., Mowlavi, N., et al.\ 2012, Astrophysics and Space Science, 341, 207 
\bibitem[Eyer et al.(2013)]{Eyeretal2013} Eyer, L., Holl, B., 
Pourbaix, D., et al.\ 2013, Central European Astrophysical Bulletin, 37, 
115 
\bibitem[Mignard et al.(2008)]{Mignardetal2008} Mignard, F., 
Bailer-Jones, C., Bastian, U., et al.\ 2008, IAU Symposium, 248, 224 
\bibitem[Nienartowicz et al.(2014)]{Nienartowiczetal2014} Nienartowicz, K., 
Ord{\'o}{\~n}ez Blanco, D., Guy, L., et al.\ 2014, arXiv:1411.5943
\bibitem[Rimoldini et al.(2012)]{Rimoldinietal2012} Rimoldini, L., 
Dubath, P., S{\"u}veges, M., et al.\ 2012, MNRAS, 427, 2917 
\bibitem[Tingley(2011)]{Tingley2011} Tingley, B.\ 2011, A\&A, 529, AA6
\bibitem[Varadi et al.(2011)]{Varadi2011} Varadi, M., Eyer, L., Jordan, S., \& Koester, D.\ 2011, EAS Publications Series, 45, 167 

\end{thebibliography}
\end{document}